\documentclass[twocolumn]{article}

\usepackage{preprint}
\usepackage{amsmath,amssymb,amsfonts}
\usepackage[numbers,square]{natbib}
\usepackage[utf8]{inputenc}
\usepackage[T1]{fontenc}
\usepackage{graphicx}
\usepackage{booktabs}
\usepackage{multirow}
\usepackage{array}
\usepackage{microtype}
\usepackage{xcolor}
\usepackage{float}
\usepackage{cuted}
\usepackage[hypcap=false]{caption}
\usepackage[colorlinks=true,linkcolor=purple,urlcolor=blue,citecolor=cyan]{hyperref}

\newcommand{\best}[1]{\textbf{#1}}
\newcommand{\second}[1]{\underline{#1}}
\newcommand{\sg}{\operatorname{sg}}
\fancypagestyle{firstpage}{%
\fancyhf{}
\cfoot{\thepage}
}

\title{Efficient JPEG Restoration in the Wavelet Domain via Mean Flows}

\usepackage{authblk}

\author[1]{\c{S}tefan-Alexandru Asandei \textsuperscript{*}}
\author[2]{Mihai-Alexandru Radu \textsuperscript{*}}

\affil[1]{Independent Researcher}
\affil[2]{Faculty of Physics, "Alexandru Ioan Cuza" University of Ia\c{s}i}

\begin{document}
\setlength{\footskip}{12pt}

\twocolumn[{%
\begin{@twocolumnfalse}
\maketitle
\begin{abstract}
Latest JPEG restoration systems achieve strong quality with large models, yet often remain too slow and expensive for efficient on-device deployment.  We present a 65M-parameter generative restorer that attains the lowest LPIPS at QF 10 and 20 on LIVE-1, Urban100, and DIV2K-val while sustaining 8.05 images/s at $1024\times1024$ on a single RTX 3090, roughly $4.9\times$ the reported throughput of one-step SODiff at one-twentieth of its parameters.  Trained from scratch, the model replaces the learned VAE encoder-decoder with an exactly invertible two-level Haar transform, predicts a clean wavelet-domain residual through a rank-enhanced linear-attention DiT that estimates compression severity internally, and is optimized with an improved MeanFlow objective that enables inference in one or two network evaluations without distillation.  Large pretrained priors remain stronger under severe compression (QF 5), whereas our model prioritizes throughput for deployment-constrained restoration.
\end{abstract}
\vspace{0.15cm}
\end{@twocolumnfalse}
}]
\thispagestyle{firstpage}

\begingroup
\renewcommand{\thefootnote}{}
\footnotetext{\hspace{-1.5em}$^*$Correspondence:\\
\hspace*{1.5em}\c{S}. Asandei: \texttt{asandei.stefanel@gmail.com}\\
\hspace*{1.5em}M. Radu: \texttt{mihai.alexradu17@gmail.com}}
\endgroup

\section{Introduction}

JPEG remains the dominant image format because its blockwise transform coding and quantization offer an effective rate-distortion trade-off.  At aggressive compression levels, however, the quantized high-frequency coefficients are unrecoverable at decode time, producing blocking, ringing, color discontinuities, and lost texture~\cite{wallace1991jpeg}.  Restoration is therefore inherently ill-posed: a method must suppress compression-induced structure without hallucinating detail that the input does not support.

Feed-forward restoration networks address this problem efficiently, but their regression objectives tend to favor over-smoothed estimates.  Recent generative approaches instead repurpose pretrained text-to-image diffusion models as image priors.  DiffBIR and SUPIR demonstrate the strength of such priors for blind restoration~\cite{lin2024diffbir,yu2024supir}, and SODiff specializes this line of work to JPEG restoration, reducing denoising to a single step~\cite{yang2025sodiff}.  The computational burden of SODiff, however, does not stem from a long sampling trajectory.  Inference instead retains the full Stable Diffusion 2.1 latent pipeline, including the denoising UNet, VAE encoding and decoding, and task-specific semantic conditioning.  This design achieves excellent perceptual quality, particularly under severe compression, but it constitutes a difficult operating point for memory and latency-constrained deployment.

\begin{figure}[t]
\centering
\includegraphics[width=\linewidth]{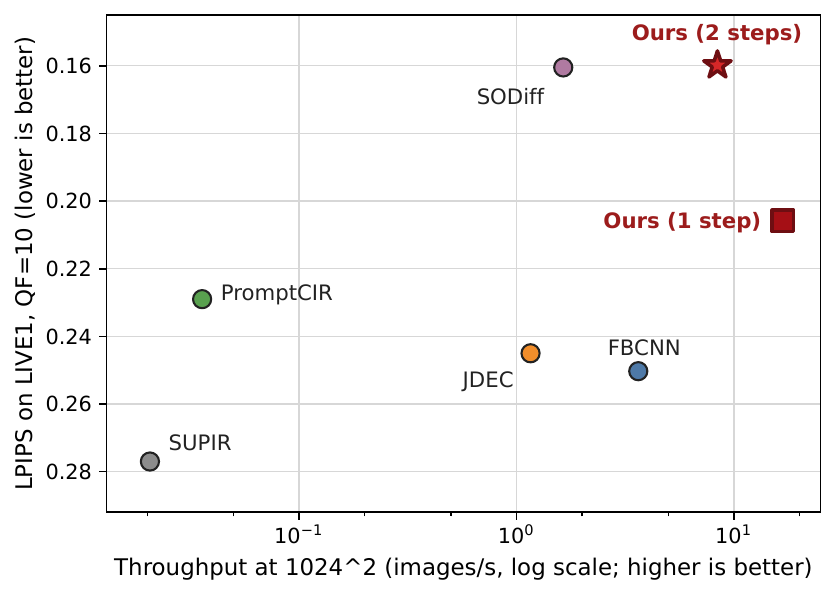}
\caption{The quality-throughput frontier on LIVE-1 at QF 10 for $1024\times1024$ inputs.  Two-update MeanFlow approaches SODiff's LPIPS while moving to a substantially higher-throughput regime; one-update MeanFlow exposes a faster operating point.  Throughput uses a logarithmic axis.}
\label{fig:pareto}
\end{figure}

This work investigates whether generative JPEG restoration can occupy a markedly different point on the quality-efficiency frontier.  We train a 65.32M-parameter model from scratch, with no text encoder, no pretrained image generator, and no VAE.  The model operates on an exactly invertible two-level Haar representation, within which a rank-enhanced linear-attention DiT predicts a clean residual quantity, an improved MeanFlow objective then trains the induced average velocity in velocity space.  This construction couples the direct clean-image prediction of JiT and pixel MeanFlow~\cite{li2025jit,lu2026pixelmeanflow} with the large-interval sampling of MeanFlow~\cite{geng2025meanflow,geng2025improved}.  Two network evaluations yield the best overall operating point in our experiments, and a single-evaluation variant is available when throughput is the priority.

We do not claim uniformly superior restoration.  A compact model trained from scratch cannot inherit the semantic and texture knowledge of a billion-parameter text-to-image model, and this deficit is visible in DISTS and no-reference quality metrics.  The model nevertheless matches or improves LPIPS at QF 10 and 20 on LIVE-1, Urban100, and DIV2K-val while sustaining 8.05 images/s at $1024\times1024$ on a single RTX 3090.  We therefore position the result as a practical Pareto point rather than a replacement for methods whose primary objective is maximal generative quality.

Our contributions are:
\begin{itemize}
\item a compact, latent-free conditional MeanFlow formulation for JPEG restoration based on clean-residual prediction in a two-level Haar domain;
\item an efficient restoration backbone that integrates RELA into a SANA-style linear DiT and estimates unknown JPEG severity internally;
\item an exact segmented JVP implementation that reduces MeanFlow training memory from 15.75 to 6.16 GiB in a controlled comparison; and
\item a quality-efficiency study spanning three datasets, six inference configurations, resolution scaling, and deployment-oriented performance.
\end{itemize}

\section{Related Work}

\subsection{JPEG artifact removal}

Blind JPEG restoration must cope with unknown compression severity.  FBCNN predicts a quality factor and conditions restoration strength on it, providing a strong and flexible convolutional baseline~\cite{jiang2021fbcnn}.  JDEC reasons over continuous cosine coefficients and the JPEG decoding pipeline~\cite{han2024jdec}, while PromptCIR learns degradation prompts for blind compressed-image restoration~\cite{li2024promptcir}.  These methods dispense with a large generative backbone and generally preserve image structure well. However, severe quantization, admits multiple plausible high-frequency reconstructions, which a deterministic regression model may average into over-smoothed output.

\subsection{Generative restoration priors}

Latent diffusion models shift generation into a learned autoencoder representation, rendering large-scale text-to-image pretraining tractable~\cite{rombach2022ldm}.  Restoration methods then reuse the resulting natural-image prior: DiffBIR couples a restoration module to a pretrained diffusion model, and SUPIR scales both the prior and the conditioning for photo-realistic restoration~\cite{lin2024diffbir,yu2024supir}.  SODiff is the closest point of comparison to our setting, performing one-step JPEG restoration with Stable Diffusion 2.1, a semantically aligned image-prompt extractor, and a QF-aware timestep predictor~\cite{yang2025sodiff,stable_diffusion_21}.  Although it avoids an online vision-language model at inference, it retains the pretrained UNet and the frozen VAE encoder-decoder.  Our model makes the opposite trade-off, forgoing this strong semantic prior in favor of a much smaller direct image-to-image system.

\subsection{Fast flows and efficient transformers}

Rectified Flow learns a transport ODE and typically draws samples through numerical integration~\cite{liu2023rectifiedflow}.  MeanFlow instead models the average velocity over a time interval, enabling the interval to be traversed in a single update without teacher distillation~\cite{geng2025meanflow}; its improved variant recasts the objective as an instantaneous-velocity loss parameterized by an average-velocity predictor, which stabilizes training~\cite{geng2025improved}.  JiT advocates direct clean-image prediction~\cite{li2025jit}, and pixel MeanFlow reconciles this prediction space with MeanFlow's velocity-space loss for latent-free generation~\cite{lu2026pixelmeanflow}.  We adapt this combination to conditional residual restoration rather than unconditional generation.

On the architectural side, DiT establishes the scalability of transformers for diffusion~\cite{peebles2023dit}, and SANA replaces quadratic self-attention with linear attention for efficient high-resolution synthesis~\cite{xie2024sana}.  Vanilla linear attention, however, can suffer a collapse in representation rank, which is particularly detrimental to local restoration.  RELA remedies this with a lightweight depthwise-convolution branch that restores local feature diversity while preserving linear scaling in the token count~\cite{ai2025rela}; we adopt the RELA operator rather than the full LAformer architecture.  Restormer~\cite{zamir2022restormer} and the ADM UNet~\cite{dhariwal2021adm} provide further context on restoration and generative backbones.

\section{Method}
\label{sec:method}

\subsection{Problem and wavelet representation}

Let $y\in[0,1]^{3\times H\times W}$ denote a clean image and $c=\mathcal{J}_{q}(y)$ its JPEG-compressed observation under an unknown quality factor $q$.  We seek a stochastic mapping $G(c,\epsilon)$ that suppresses compression artifacts while remaining faithful to the content observed in $c$.

Rather than pixel space or the latent space of a learned VAE, we operate in the domain of a two-level orthonormal Haar wavelet-packet transform $\mathcal{H}^{2}$~\cite{mallat1989wavelet}.  One transform level maps each $2\times2$ neighborhood $(a,b;c,d)$ to an approximation coefficient and three detail coefficients,

\vspace{-1em}
\begin{equation}
\begin{aligned}
\ell &= \tfrac{a+b+c+d}{2}, & h &= \tfrac{a-b+c-d}{2}, \\
v &= \tfrac{a+b-c-d}{2}, & g &= \tfrac{a-b-c+d}{2}.
\end{aligned}
\label{eq:haar}
\end{equation}
which separate the local average from horizontal, vertical, and diagonal variation.  Orthonormality makes the mapping exactly invertible and energy-preserving.  We apply the same transform to every first-level subband, not only the approximation band, giving the tensor-shape sequence $3\times H\times W\rightarrow12\times H/2\times W/2\rightarrow 48\times H/4\times W/4$.  The transform is parameter-free and adds negligible endpoint cost, yet it gives the transformer a compact, frequency-organized grid without the reconstruction error or decoder latency of a learned autoencoder.

Writing $h_c=\mathcal{H}^{2}(c)$ for the encoded condition, we model the scaled residual
\begin{equation}
x=s\big(\mathcal{H}^{2}(y)-h_c\big),\qquad s=16
\label{eq:residual}
\end{equation}
rather than the full clean coefficient map.  A predicted residual $\hat{x}$ is decoded as $\hat{y}=\mathcal{H}^{-2}(h_c+\hat{x}/s)$.  This parameterization places the observed low-frequency content on a fixed skip path and concentrates model capacity on the components that quantization has corrupted or removed.

\subsection{Conditional pixel MeanFlow}

We adopt the time convention $t=0$ for data and $t=1$ for noise.  Given $\epsilon\sim\mathcal{N}(0,I)$, the conditional flow path and its velocity are
\begin{equation}
z_t=(1-t)x+t\epsilon, \qquad v_c=\epsilon-x.
\label{eq:path}
\end{equation}
Rectified Flow regresses the instantaneous velocity $v_c$ and must integrate it over several numerical steps.  MeanFlow instead learns the \emph{average} velocity over an interval $[r,t]$,
\begin{equation}
u(z_t,r,t)=\frac{1}{t-r}\int_r^t v(z_\tau,\tau)\,d\tau,
\label{eq:avgvel}
\end{equation}
so that the whole interval is traversed by the single update $z_r=z_t-(t-r)\,u(z_t,r,t)$.  Differentiating $(t-r)\,u(z_t,r,t)$ with respect to $t$ gives the MeanFlow identity
\begin{equation}
v(z_t,t)=u(z_t,r,t)+(t-r)\frac{d\,u(z_t,r,t)}{dt},
\label{eq:identity}
\end{equation}
where the total derivative is taken along the flow trajectory.  It is this identity that makes one- or two-evaluation restoration trainable without teacher distillation.

Following the clean-data parameterization of JiT and pixel MeanFlow~\cite{li2025jit,lu2026pixelmeanflow}, the network predicts the clean residual endpoint directly and converts it to an average velocity:
\begin{equation}
x_\theta=f_\theta\big([z_t,h_c],t-r\big),\qquad
u_\theta=\frac{z_t-x_\theta}{\max(t,t_{\min})},
\label{eq:xpred}
\end{equation}
with $t_{\min}=0.05$.  The network is conditioned on the interval length $t-r$ rather than on two absolute time embeddings, which keeps the direct prediction near the structured residual manifold while the regression target remains a velocity.

We train with improved MeanFlow~\cite{geng2025improved}, which instantiates the identity~\eqref{eq:identity} using the network's own prediction.  The corrected instantaneous velocity $V_\theta=u_\theta+(t-r)\,\sg\big(d\,u_\theta/dt\big)$ is formed by evaluating the total derivative of $u_\theta$ along the flow path as an exact Jacobian-vector product (JVP), and is regressed onto $v_c$ with adaptive weighting:
\begin{equation}
d=\lVert V_\theta-v_c\rVert_2^2, \qquad
\mathcal{L}_{\mathrm{MF}}=(\sg(d)+0.01)^{-1}d.
\label{eq:mfloss}
\end{equation}
Gradients flow through the average-velocity prediction but are stopped through the JVP correction, as prescribed by improved MeanFlow.  Section~\ref{sec:jvp} describes how we evaluate this exact derivative with lower training memory.

Because $x_\theta$ decodes directly into an image estimate, we additionally apply LPIPS~\cite{zhang2018lpips} to $\mathcal{H}^{-2}(h_c+x_\theta/s)$ when $t<0.8$.  The complete objective augments the MeanFlow loss with this perceptual term and a severity regression on the head described in the next subsection:
\begin{equation}
\mathcal{L}=\mathcal{L}_{\mathrm{MF}}+\lambda_p\mathcal{L}_{\mathrm{LPIPS}}
+\lambda_q\left|\hat{d}_q-(1-q/100)\right|,
\end{equation}
with $\lambda_p=0.4$ in the final stage and $\lambda_q=0.1$.  Sampling requires no iterative ODE solver and costs one network evaluation per requested interval:
\begin{equation}
z_r=z_t-(t-r)u_\theta(z_t,r,t).
\label{eq:sampling}
\end{equation}
We initialize $z_1=\epsilon$ and apply Eq.~\eqref{eq:sampling} over one or two uniform intervals.

\subsection{Rank-enhanced linear DiT}

The backbone follows the SANA linear-DiT layout~\cite{xie2024sana}: 16 transformer blocks with hidden width 576, 18 heads of dimension 32, and an MLP ratio of 2.5.  A $2\times2$ patch embedding in the two-level Haar domain yields a token grid of $H/8\times W/8$.  The noisy state and the degraded condition are concatenated channel-wise, following the direct image-conditioning strategy of Palette~\cite{saharia2022palette}, and cross-attention is disabled.

For $N$ tokens, vanilla linear attention reassociates the attention product as $\phi(Q)(\phi(K)^{T}V)$ and therefore scales linearly in $N$, but the resulting attention map has rank at most the head channel dimension.  We insert the RELA branch into every block:
\begin{equation}
\operatorname{RELA}(Q,K,V)=
\frac{\phi(Q)(\phi(K)^TV)}{\phi(Q)(\phi(K)^T\mathbf{1})}
+\operatorname{DWConv}_{5\times5}(V),
\end{equation}
where $\phi$ is ReLU in our implementation.  The depthwise convolution contributes a local, potentially full-rank path at small parameter and runtime cost.

The JPEG severity is unknown at inference and is estimated internally.  After the first three DiT blocks, global average pooling and a small MLP predict $\hat d_q\in[0,1]$, whose embedding is added to the interval embedding for the remaining 13 blocks.  This design is inspired by FBCNN's explicit quality prediction, except that the predicted value modulates the same transformer rather than a separate decoder~\cite{jiang2021fbcnn}.

\subsection{Memory-efficient exact JVP}
\label{sec:jvp}

MeanFlow training requires differentiating the network output along a tangent direction.  A monolithic \texttt{torch.func.jvp} retains tangent activations across all blocks and is memory-intensive at restoration resolutions.  Inspired by the segmented network treatment in rCM~\cite{rcm2025code}, we partition the forward pass into the patch and time embeddings, the individual transformer blocks, the quality head, and the output projection.  Each segment evaluates an exact JVP and detaches only the outgoing tangent before the next segment, while the primal path remains fully differentiable for the training loss.  This leaves both the tangent and the loss numerically unchanged, but releases intermediate tangent graphs earlier.  We do not use the custom FlashAttention JVP kernel~\cite{jvpflash2025code}, because our RELA operator is not scaled dot-product attention.

\section{Experiments}
\label{sec:experiments}

\subsection{Datasets, training, and evaluation}

\paragraph{Training data.}
We train on DF2K (3,450 images)~\cite{agustsson2017div2k} and 10,000 images drawn from two LSDIR shards~\cite{li2023lsdir}.  The final checkpoint comprises approximately 101k optimizer updates: 220 epochs on the combined 13,450-image pool with random $128\times128$ crops at batch size 32, followed by seven epochs on a 5,000-image LSDIR shard with $368\times368$ crops at batch size 4.  The first stage samples the QF uniformly from 5-40, the large-crop continuation concentrates on QF 5-20.  Random horizontal and vertical flips provide the only augmentation.

All models are trained from scratch on a single RTX 3090 with AdamW~\cite{loshchilov2019adamw} ($\beta=(0.9,0.95)$, no weight decay) and gradient-norm clipping at 5.  The base learning rate is $5\times10^{-5}$ after warm-up and is reduced to $2\times10^{-5}$ for the large-crop continuation, the LPIPS weight is 0.2 in the base stage and 0.4 in the final stage.  Explorative modelling with $K=3$ candidates~\cite{gladstone2026xm} accelerated early convergence by roughly 20\% in preliminary rectified-flow runs, the reported MeanFlow training uses $K=1$.

The two stages serve complementary purposes.  Small crops keep the first 100k updates affordable while exposing the model to many compression realizations, and the $368\times368$ continuation adapts the positional interpolation and attention blocks to longer-range structure while shifting weight toward the QF range used in the main benchmark.  A single final checkpoint is selected and used for every dataset and QF, with no dataset-specific tuning.

\paragraph{Benchmarks and metrics.}
We evaluate full images from LIVE-1~\cite{sheikh2005live1}, Urban100~\cite{huang2015urban100}, and DIV2K-val~\cite{agustsson2017div2k} at QF $\{5,10,20\}$.  Full-reference metrics are LPIPS~\cite{zhang2018lpips} and DISTS~\cite{ding2020dists}; no-reference metrics are MUSIQ~\cite{ke2021musiq}, MANIQA~\cite{yang2022maniqa}, and CLIPIQA~\cite{wang2023clipiqa}.  Quality figures for competing methods are taken from SODiff under an identical full-image protocol~\cite{yang2025sodiff}.  Unless stated otherwise, we report the two-update MeanFlow schedule with the random seed fixed to zero.  The evaluation QF is used only to synthesize the JPEG input and is never given to the network, which observes the decoded RGB image and estimates degradation severity internally.  Full-image inference avoids the overlap, blending, and crop-selection effects introduced by tiled evaluation.

\subsection{Main results}

Table~\ref{tab:quality} presents the complete comparison.  Our model matches SODiff on LIVE-1 at QF 10 and attains the lowest LPIPS at QF 10 and 20 on the other dataset-QF pairs.  The advantage is largest at QF 20, where LPIPS decreases relative to SODiff by 25.9\% on LIVE-1, 36.4\% on Urban100, and 23.2\% on DIV2K-val.  At QF 10 the margins are considerably smaller (0.4\%, 6.5\%, and 2.1\%, respectively), and these figures are therefore best interpreted as competitive quality at a substantially different compute point rather than as broad quality dominance.

This pattern across compression levels is consistent with the model design.  At QF 20, edge and texture evidence survives in the decoded image and the wavelet-domain residual model refines it effectively.  At QF 5, much of that evidence has been quantized away, and SODiff improves LIVE-1 LPIPS from our 0.2864 to 0.2229, with comparable gaps on Urban100 and DIV2K-val: a pretrained semantic prior is most valuable precisely in this low-information regime.

\vspace{2.5cm}

\begin{strip}
\centering
\begin{minipage}{\textwidth}
\centering
\scriptsize
\setlength{\tabcolsep}{2.5pt}
\renewcommand{\arraystretch}{0.85}
\begin{tabular*}{\textwidth}{@{\extracolsep{\fill}}l|ccc|ccc|ccc|ccc|ccc@{}}
\toprule
\multicolumn{16}{c}{\textbf{LIVE-1}}\\
\midrule
\multirow{2}{*}{Method} & \multicolumn{3}{c|}{LPIPS $\downarrow$} & \multicolumn{3}{c|}{DISTS $\downarrow$} & \multicolumn{3}{c|}{MUSIQ $\uparrow$} & \multicolumn{3}{c|}{MANIQA $\uparrow$} & \multicolumn{3}{c}{CLIPIQA $\uparrow$}\\
& QF5&QF10&QF20&QF5&QF10&QF20&QF5&QF10&QF20&QF5&QF10&QF20&QF5&QF10&QF20\\
\midrule
JPEG & 0.4384&0.3013&0.1799&0.3242&0.2387&0.1653&40.33&53.88&64.12&0.2294&0.3509&0.4411&0.1716&0.2737&0.5542\\
FBCNN & 0.3736&0.2503&0.1583&0.2353&0.1785&0.1319&\second{63.56}&71.00&73.96&\second{0.3425}&0.4207&0.4551&0.2763&0.4767&0.5535\\
JDEC & 0.4113&0.2450&0.1555&0.2364&0.1740&0.1282&55.66&70.80&73.81&0.2002&0.4065&0.4433&0.1539&0.4811&0.5512\\
PromptCIR & 0.3797&0.2290&0.1450&0.2334&0.1658&0.1223&60.34&\second{72.39}&\best{74.12}&0.2790&0.4500&0.4713&0.2655&0.5176&0.5847\\
DiffBIR* & 0.3509&0.2160&0.1500&\second{0.2035}&\second{0.1319}&\second{0.0988}&58.09&67.38&71.08&0.2812&0.3789&0.4371&0.3776&0.5789&0.6814\\
SUPIR & 0.4856&0.2770&0.1683&0.2720&0.1558&0.1121&52.69&68.77&73.02&0.3229&\second{0.5183}&\best{0.6237}&0.3149&0.6115&\second{0.7364}\\
SODiff & \best{0.2229}&\second{0.1605}&\second{0.1237}&\best{0.1173}&\best{0.0938}&\best{0.0763}&\best{72.88}&\best{73.84}&\second{74.11}&\best{0.4957}&\best{0.5192}&\second{0.5272}&\best{0.7087}&\best{0.7323}&\best{0.7587}\\
\textbf{Ours} & \second{0.2864}&\best{0.1599}&\best{0.0917}&0.2070&0.1443&0.1008&63.37&68.99&71.90&0.2602&0.3325&0.3751&\second{0.6083}&\second{0.6846}&0.7358\\
\midrule
\multicolumn{16}{c}{\textbf{Urban100}}\\
\midrule
JPEG & 0.3481&0.2254&0.1244&0.2834&0.2145&0.1521&50.46&60.87&67.60&0.3656&0.4401&0.4967&0.2806&0.3517&0.5343\\
FBCNN & 0.2341&0.1462&0.0896&0.2162&0.1648&0.1249&69.03&\second{72.55}&\second{73.39}&0.4263&0.5033&0.5288&0.3800&0.5014&0.5437\\
JDEC & 0.2794&0.1382&0.0846&0.2309&0.1570&0.1175&62.97&72.52&73.30&0.3386&0.5001&0.5230&0.2518&0.4959&0.5369\\
PromptCIR & 0.2389&0.1183&\second{0.0739}&0.2037&0.1431&0.1083&66.08&\best{73.01}&\best{73.47}&0.3946&0.5380&0.5489&0.3619&0.5337&0.5662\\
DiffBIR* & \second{0.2018}&0.1344&0.1005&\second{0.1657}&\second{0.1207}&\second{0.0939}&69.63&71.77&72.51&0.4285&0.4813&0.5105&0.5470&0.5966&0.6306\\
SUPIR & 0.3279&0.2489&0.2125&0.2018&0.1659&0.1518&\second{69.94}&72.37&73.01&\second{0.5546}&\best{0.5995}&\best{0.6105}&0.5536&0.6178&0.6397\\
SODiff & \best{0.1579}&\second{0.1098}&0.0846&\best{0.1196}&\best{0.0914}&\best{0.0734}&\best{71.33}&72.23&72.63&\best{0.5598}&\second{0.5451}&\second{0.5561}&\second{0.6392}&\second{0.6558}&\second{0.6733}\\
\textbf{Ours} & 0.2045&\best{0.1027}&\best{0.0538}&0.2560&0.1880&0.1368&65.31&68.92&70.24&0.3598&0.4258&0.4682&\best{0.6428}&\best{0.6939}&\best{0.7172}\\
\midrule
\multicolumn{16}{c}{\textbf{DIV2K-val}}\\
\midrule
JPEG & 0.3459&0.3234&0.2072&0.2570&0.2255&0.1465&25.95&47.53&57.45&0.2570&0.3120&0.3557&0.2595&0.3303&0.5072\\
FBCNN & 0.3445&0.2448&0.1733&0.2078&0.1581&0.1168&56.52&61.79&65.20&0.3025&0.3593&0.3775&0.3004&0.4561&0.5221\\
JDEC & 0.3811&0.2313&0.1565&0.2234&0.1574&0.1152&53.88&\best{67.48}&\best{69.90}&0.2118&0.3689&\second{0.3927}&0.1841&0.4675&0.5319\\
PromptCIR & 0.3549&0.2240&0.1581&0.2067&0.1459&\second{0.1061}&52.21&62.63&65.62&0.2705&\second{0.3758}&0.3871&0.3041&0.4956&0.5483\\
DiffBIR* & \second{0.2788}&0.1953&0.1542&\second{0.1533}&\second{0.1072}&0.0856&\second{60.21}&65.22&67.06&0.3220&0.3754&\best{0.4033}&0.4975&0.5912&0.6355\\
SUPIR & 0.4372&0.3121&0.2295&0.2148&0.1410&0.1161&54.07&61.93&64.87&\second{0.3438}&0.3570&0.3723&0.4219&0.5186&0.5535\\
SODiff & \best{0.2425}&\second{0.1732}&\second{0.1295}&\best{0.1126}&\best{0.0816}&\best{0.0622}&\best{64.42}&\second{65.90}&\second{66.49}&\best{0.3733}&\best{0.3924}&\second{0.3984}&\second{0.5851}&\second{0.6193}&\second{0.6398}\\
\textbf{Ours} & 0.2821&\best{0.1695}&\best{0.0995}&0.1800&0.1209&\second{0.0828}&57.36&62.30&64.62&0.2494&0.3103&0.3427&\best{0.6048}&\best{0.6558}&\best{0.6820}\\
\bottomrule
\end{tabular*}
\captionof{table}{Full-image quantitative comparison at JPEG quality factors 5, 10, and 20.  Best results are bold and second-best results are underlined.  Comparison values are from SODiff~\cite{yang2025sodiff}; DiffBIR* denotes its JPEG-restoration retraining.  Our two-update MeanFlow is evaluated with the same datasets and protocol.}
\label{tab:quality}
\end{minipage}
\end{strip}

The remaining metrics make the trade-off explicit.  SODiff dominates DISTS, and the large pretrained restorers are generally stronger on MUSIQ and MANIQA.  Our weakest relative result is MANIQA, where a local artifact prior cannot substitute for object- and texture-specific knowledge.

In contrast, CLIPIQA is strongest on Urban100 and DIV2K-val at every tested QF, suggesting that semantic-image alignment can remain plausible even when fine perceptual statistics depart from those favored by DISTS or MANIQA.

\subsection{Flow objective ablation}

Table~\ref{tab:samplers} controls for the backbone and training budget, comparing one- and two-update MeanFlow schedules with Euler and Heun integration of the rectified-flow baseline.  MeanFlow is a learned interval update rather than a numerical sampler: ``uniform-2'' applies Eq.~\eqref{eq:sampling} over $[1,0.5]$ and $[0.5,0]$, while ``direct-1'' traverses $[1,0]$ in a single step.  The two-update schedule achieves the best LPIPS at every QF using two network evaluations, against 20-100 for the RF solvers.  This advantage is consistent with the training signal: MeanFlow's endpoint prediction is decoded directly for the LPIPS loss, whereas the RF model evaluates the loss on an endpoint estimated from an intermediate state.  RF nevertheless remains stronger on DISTS and several no-reference measures, so the result is specific to the quality criterion rather than a uniform inference win.

The second MeanFlow update is most beneficial when the JPEG input still carries recoverable detail: relative to one update, it reduces LPIPS by 22.3\% at QF 10 and 31.5\% at QF 20, but by only 6.6\% at QF 5.  The second interval thus appears to refine image-supported structure more effectively than it invents detail removed by severe quantization.  Conversely, increasing Euler integration from 20 to 50 steps worsens LPIPS at QF 10 and 20, and Heun is not consistently better than Euler.  Numerical integration error is therefore not the sole limiting factor, the learned velocity field and the perceptual objective also shape the endpoint.  We adopt the uniform-2 MeanFlow schedule for the main comparison and retain direct-1 as the throughput-oriented option in Fig.~\ref{fig:pareto}.

\begin{strip}
\centering
\begin{minipage}{\textwidth}
\centering
\footnotesize
\setlength{\tabcolsep}{3.5pt}
\renewcommand{\arraystretch}{0.85}

\begin{tabular}{lllrrrrrrr}
\toprule
Objective & Sampler & Integration Steps & NFE & QF
& LPIPS$\downarrow$
& DISTS$\downarrow$
& MUSIQ$\uparrow$
& MANIQA$\uparrow$
& CLIPIQA$\uparrow$ \\
\midrule

\multirow{6}{*}{Mean Flow}
& \multirow{6}{*}{Uniform}
& \multirow{3}{*}{2 steps}
& \multirow{3}{*}{2}
& 5  & \best{0.286388} & 0.206975 & 63.3710 & 0.260224 & 0.608269 \\
&
&
&
& 10 & \best{0.159868} & 0.144288 & 68.9947 & 0.332499 & 0.684581 \\
&
&
&
& 20 & \best{0.091668} & 0.100836 & 71.9031 & 0.375095 & 0.735817 \\
\cmidrule(lr){3-10}
&
&
\multirow{3}{*}{1 step}
& \multirow{3}{*}{1}
& 5  & \second{0.306523} & 0.200740 & 60.9723 & 0.234335 & 0.554503 \\
&
&
&
& 10 & 0.205874 & 0.148648 & 67.6719 & 0.319600 & 0.639117 \\
&
&
&
& 20 & 0.133789 & 0.109228 & 71.8606 & 0.383767 & 0.696279 \\

\midrule

\multirow{12}{*}{Rectified Flow}
& \multirow{6}{*}{Euler}
& \multirow{3}{*}{20 steps}
& \multirow{3}{*}{20}
& 5  & 0.322365 & \second{0.198518} & 65.9203 & 0.364685 & 0.634485 \\
&
&
&
& 10 & \second{0.177425} & \best{0.134750} & 70.3500 & 0.397286 & 0.714172 \\
&
&
&
& 20 & \second{0.098188} & \best{0.090350} & \best{72.7822} & 0.430171 & 0.737860 \\
\cmidrule(lr){3-10}
&
&
\multirow{3}{*}{50 steps}
& \multirow{3}{*}{50}
& 5  & 0.311847 & \best{0.197340} & 67.5549 & 0.360032 & \second{0.640977} \\
&
&
&
& 10 & 0.186408 & \second{0.139035} & 70.6819 & 0.405368 & \best{0.714576} \\
&
&
&
& 20 & 0.105492 & \second{0.094806} & 72.6318 & 0.442486 & 0.742111 \\

\cmidrule(lr){2-10}

&
\multirow{6}{*}{Heun}
& \multirow{3}{*}{20 steps}
& \multirow{3}{*}{40}
& 5  & 0.334665 & 0.202531 & \second{67.9828} & \best{0.381495} & 0.638597 \\
&
&
&
& 10 & 0.202867 & 0.144393 & \best{70.9587} & \best{0.424092} & 0.712926 \\
&
&
&
& 20 & 0.113689 & 0.098517 & \second{72.7513} & \best{0.459066} & \second{0.743271} \\
\cmidrule(lr){3-10}
&
&
\multirow{3}{*}{50 steps}
& \multirow{3}{*}{100}
& 5  & 0.316402 & 0.198977 & \best{68.4464} & \second{0.367442} & \best{0.642883} \\
&
&
&
& 10 & 0.198745 & 0.143537 & \second{70.8858} & \second{0.416914} & \second{0.714408} \\
&
&
&
& 20 & 0.113569 & 0.098914 & 72.5574 & \second{0.454989} & \best{0.744678} \\

\bottomrule
\end{tabular}

\captionof{table}{ Flow-objective and inference-update ablation on LIVE-1.  Mean Flow and Rectified Flow use the same 65M-parameter RELA DiT family and comparable training budgets. Best results at each QF are bold and second-best results are underlined.  }
\label{tab:samplers}

\end{minipage}
\end{strip}

\subsection{Efficiency and resolution scaling}
\enlargethispage{3\baselineskip}

We compare complete restoration systems at $1024\times1024$ with batch size one (Table~\ref{tab:system_efficiency}).  The final two-update model reaches 8.05 images/s.  Against the externally reported one-step SODiff figure, this is a cross-paper $4.9\times$ comparison rather than a controlled speedup.  This reflects the architecture: our 65.32M-parameter model acts directly in an invertible wavelet representation, while SODiff reports 1,288M parameters and uses VAE encoding and decoding around its one-step UNet.

\vspace{-0.5\baselineskip}
\begin{table}[H]
\centering
\scriptsize
\setlength{\tabcolsep}{3pt}
\renewcommand{\arraystretch}{0.85}
\caption{System-level efficiency for batch-one $1024\times1024$ restoration.}
\label{tab:system_efficiency}
\resizebox{\columnwidth}{!}{%
\begin{tabular}{lrrrr}
\toprule
Method & \shortstack{Parameters\\(M)} & NFE & \shortstack{Latency\\(s)} & \shortstack{Throughput\\(img/s)}\\
\midrule
FBCNN &70.10&1&0.275&3.63\\
JDEC &38.90&1&0.860&1.16\\
PromptCIR &34.75&64&27.97&0.0358\\
SUPIR &4490&50&48.66&0.0206\\
SODiff &1288&1&0.610&1.64\\
\textbf{Ours} &65.32&2&0.1242&8.05\\
\bottomrule
\end{tabular}}
\end{table}

This compares deployed systems rather than isolated attention kernels.  NFE counts restoration-network evaluations, while latency includes encoding and decoding.  Our 0.1242 s latency is $2.2\times$ lower than FBCNN and $4.9\times$ lower than SODiff.  The controlled same-GPU study below isolates backbone scaling. Parameter count reinforces this difference.  Our model is approximately $19.7\times$ smaller than SODiff and $68.7\times$ smaller than SUPIR, while SODiff remains slower despite one denoising evaluation because of its large UNet and VAE endpoints.  The comparison therefore reflects backbone size, representation overhead, and update count together.

This distinction matters for deployment: a one-step model is not necessarily efficient if each evaluation still passes through a large latent pipeline.  Our two-update path avoids a learned encoder, decoder, and external conditioning model, so its latency reflects the complete restoration path rather than the sampler alone.  The next experiment isolates backbone scaling under a shared wavelet representation, precision, hardware, and eager execution.

Resolution scaling is central to restoration because the input grid must be preserved.  Here an $H\times W$ image yields $N=HW/64$ tokens after the Haar and patch transforms.  Full attention constructs an $N\times N$ map, so doubling each image dimension quadruples $N$ and increases its attention term by $16\times$.  Linear attention reassociates the product to scale linearly in $N$~\cite{xie2024sana}, but constrains the effective attention-map rank to the per-head channel dimension.  This can suppress local feature diversity as the token grid grows.  RELA adds a depthwise-convolution path that is local and potentially full-rank while preserving linear token scaling~\cite{ai2025rela}.  The controlled study below compares the resulting throughput and memory trends against full attention, vanilla linear attention, and Restormer at matched capacity.

\clearpage
\twocolumn[{%
\begin{@twocolumnfalse}
\begin{minipage}{\textwidth}
\centering
\begin{minipage}[t]{0.47\textwidth}
\centering
\includegraphics[width=\linewidth]{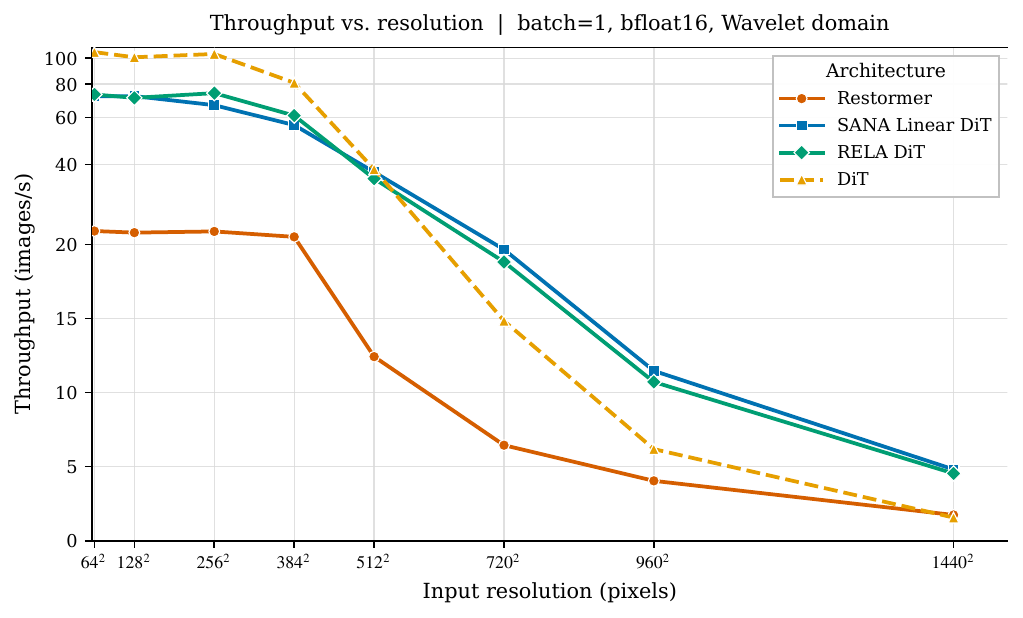}
\end{minipage}\hfill
\begin{minipage}[t]{0.47\textwidth}
\centering
\includegraphics[width=\linewidth]{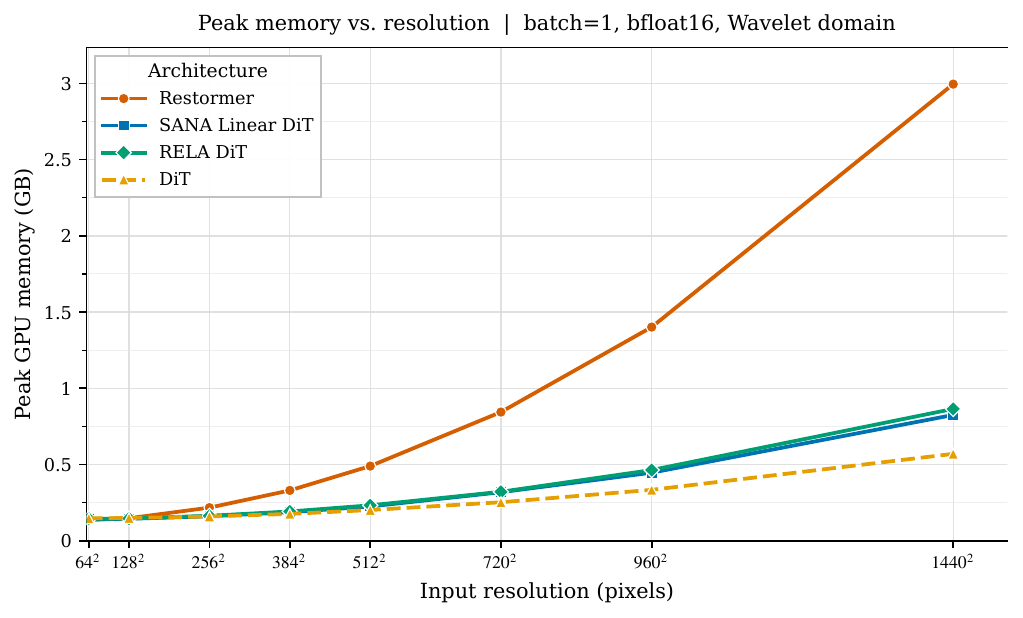}
\end{minipage}
\captionof{figure}{Controlled one-forward-pass scaling in the two-level Haar domain, batch size 1, BF16, RTX 3090.  Left: throughput.  Right: peak allocated memory.  These architecture measurements isolate backbone cost, they are not two-update end-to-end MeanFlow measurements.}
\label{fig:backbone_scaling}
\vspace{3pt}
\scriptsize
\setlength{\tabcolsep}{2pt}
\renewcommand{\arraystretch}{0.85}
\resizebox{\textwidth}{!}{%
\begin{tabular}{r|rrr|rrr|rrr|rrr}
\toprule
&\multicolumn{3}{c|}{Restormer (57.12M)}&\multicolumn{3}{c|}{SANA Linear DiT (64.34M)}&\multicolumn{3}{c|}{RELA DiT (65.32M)}&\multicolumn{3}{c}{DiT (67.47M)}\\
Resolution & ms & img/s & GiB & ms & img/s & GiB & ms & img/s & GiB & ms & img/s & GiB\\
\midrule
$64^2$   &44.4707&22.4867&0.12538&13.8961&71.9626&0.13189&13.6673&73.1673&0.13368&9.4790&105.496&0.13638\\
$128^2$  &45.0920&22.1769&0.13728&13.8650&72.1242&0.13371&14.0712&71.0673&0.13571&9.9097&100.911&0.13871\\
$256^2$  &44.6283&22.4073&0.20189&14.9908&66.7077&0.14912&13.5093&74.0231&0.15194&9.6308&103.834&0.14805\\
$384^2$  &46.7692&21.3816&0.30720&17.8032&56.1697&0.17399&16.4050&60.9571&0.17860&12.356&80.9325&0.16362\\
$512^2$  &80.4425&12.4312&0.45579&26.7253&37.4178&0.20924&28.2868&35.3522&0.21598&26.0815&38.3414&0.18736\\
$720^2$  &155.055&6.4493&0.78590&50.8455&19.6674&0.29607&53.1076&18.8297&0.29880&67.3489&14.8480&0.23488\\
$960^2$  &247.412&4.0418&1.3052&87.1724&11.4715&0.41543&93.2710&10.7214&0.43167&161.674&6.1853&0.31070\\
$1440^2$ &575.833&1.7366&2.7898&207.718&4.8142&0.76812&220.201&4.5413&0.80525&640.170&1.5621&0.53055\\
\bottomrule
\end{tabular}}
\captionof{table}{Raw one-forward-pass architecture benchmark.  All models use batch size 1, BF16, the two-level Haar representation, PyTorch eager execution, and one RTX 3090.  Resolution is square.}
\label{tab:backbone_raw}
\end{minipage}
\vspace{3pt}
\end{@twocolumnfalse}
}]

To separate backbone scaling from the number of flow evaluations, we additionally benchmark a single forward pass through four capacity-matched PyTorch architectures in BF16, without compilation or fused custom kernels.  Inputs are square ($H=W$), so ``720'' denotes $720\times720$ rather than $1280\times720$.  Figure~\ref{fig:backbone_scaling} and Table~\ref{tab:backbone_raw} show that SANA linear attention and RELA scale similarly: RELA's local depthwise branch adds a measured 5.7\% throughput cost at $1440^2$ relative to SANA linear attention while retaining the local rank-enhancing path used by the final model.  Below $512^2$, launch overheads and fixed model costs dominate and the full-attention DiT is fastest, the ordering reverses as the token grid grows, and at $1440^2$ RELA is $2.9\times$ faster than full-attention DiT.

Memory scaling tells a complementary story: Restormer reaches 2.79 GiB at $1440^2$ against 0.81 GiB for RELA, despite comparable parameter counts, while full-attention DiT uses less memory than RELA in this forward-only measurement but pays for its quadratic attention in latency.  RELA is therefore not selected because it wins every small-resolution benchmark, but because it combines favorable high-resolution scaling with an explicit local restoration path.

\subsection{Deployment optimizations}

Table~\ref{tab:deploy} separates the algorithmic gain from standard deployment optimizations.  On 16 LIVE-1 images at QF 10, replacing 20-step RF with the uniform-2 MeanFlow update provides the principal speedup and slightly improves LPIPS: batch throughput rises from 1.50 to 12.61 images/s, an $8.4\times$ gain, while LPIPS improves from 0.177425 to 0.160386.  Because this row-to-row comparison uses the same input batch and GPU, it measures the effect of the two-update formulation more directly than the cross-paper system comparison in Table~\ref{tab:system_efficiency}.

\begin{table}[H]
\centering
\footnotesize
\setlength{\tabcolsep}{3pt}
\renewcommand{\arraystretch}{0.85}
\caption{Algorithmic and implementation changes on LIVE-1 at QF 10 with a batch of 16.  Batch latency measures the complete 16-image batch.}
\label{tab:deploy}
\resizebox{\columnwidth}{!}{%
\begin{tabular}{lrrrr}
\toprule
Configuration & \shortstack{Batch latency\\(s)} & \shortstack{Throughput\\(img/s)} & \shortstack{Peak memory\\(GiB)} & LPIPS\\
\midrule
RF BF16, Euler-20 &10.6677&1.4999&2.0378&0.177425\\
MeanFlow BF16, uniform-2 &1.2689&12.6096&1.9794&0.160386\\
$+$ \texttt{torch.compile} &0.5961&26.8415&0.5107&0.160386\\
$+$ INT8 weights &0.6093&26.2585&0.4907&0.160416\\
\bottomrule
\end{tabular}}
\end{table}

Applying \texttt{torch.compile} to the inference core fuses and autotunes the static graph, we use Inductor with maximum autotuning while retaining cuDNN/ATen depthwise convolutions, whose Triton counterparts were slower on the RTX 3090.  Compilation yields a further $2.1\times$ throughput gain and reduces measured allocated memory through more effective buffer reuse.  Weight-only INT8 quantization with TorchAO then lowers allocated memory by only 0.02 GiB and slightly reduces throughput, its practical benefit is checkpoint size, which falls from 784 MB to 109 MB.  The essentially unchanged LPIPS confirms numerical robustness, but INT8 remains a standard deployment option rather than a contribution of the restoration method.

\subsection{JVP implementation}

Table~\ref{tab:jvp} evaluates the MeanFlow derivative paths on the same batch.  The segmented rCM-style implementation matches \texttt{torch.func.jvp} exactly in both tangent output and loss while reducing peak memory from 15.75 to 6.16 GiB, a 60.9\% reduction.  Runtime is effectively unchanged (286.31 versus 286.39 ms), so segmentation should be understood as a training-capacity optimization: it makes larger crops or batches feasible without altering the objective.

Finite differences illustrate why agreement of the scalar loss alone is not sufficient validation.  At $\delta=10^{-3}$, forward differences change the reported loss by only $7.15\times10^{-7}$, yet the tangent rRMSE is 0.27086 and cosine similarity drops to 0.968648.  Larger steps further distort the tangent direction, while smaller steps become increasingly sensitive to BF16 rounding.  The modest speed gains therefore do not justify injecting a biased derivative throughout training.  The FlashAttention row is only a compatibility control: because RELA is not scaled dot-product attention, its custom kernel cannot accelerate the dominant operation.  We use the exact segmented path in all final MeanFlow training.

\begin{strip}
\vspace{.1cm}
\centering
\begin{minipage}{\textwidth}
\centering
\footnotesize
\setlength{\tabcolsep}{4pt}
\renewcommand{\arraystretch}{0.85}
\begin{tabular}{lrrrrrr}
\toprule
Derivative method & ms & speedup & peak GiB & tangent rRMSE & cosine sim. & $|\Delta\mathcal L|$\\
\midrule
\texttt{torch.func.jvp} &286.31&1.00&15.75&0&1.000000&0\\
FlashAttention JVP &286.86&1.00&15.75&0&1.000000&0\\
segmented exact JVP &286.39&1.00&6.16&0&1.000000&0\\
forward FD ($10^{-3}$) &197.93&1.45&5.63&0.27086&0.968648&$7.15\!\times\!10^{-7}$\\
central FD ($10^{-3}$) &248.25&1.15&5.63&0.14659&0.989624&$2.38\!\times\!10^{-7}$\\
forward FD ($3\!\times\!10^{-3}$) &216.04&1.33&5.63&0.48775&0.928010&$2.38\!\times\!10^{-7}$\\
central FD ($3\!\times\!10^{-3}$) &243.98&1.17&5.63&0.43166&0.937275&$1.79\!\times\!10^{-7}$\\
forward FD ($10^{-2}$) &202.07&1.42&5.63&2.0637&0.710906&$1.07\!\times\!10^{-6}$\\
central FD ($10^{-2}$) &262.52&1.09&5.63&1.2534&0.807025&$8.34\!\times\!10^{-7}$\\
\bottomrule
\end{tabular}
\captionof{table}{MeanFlow JVP comparison.  rRMSE and cosine compare the tangent with
\texttt{torch.func.jvp}; $|\Delta\mathcal L|$ is the absolute loss difference.  The FlashAttention row is a compatibility control: RELA does not use its custom SDPA kernel.}
\label{tab:jvp}
\end{minipage}
\end{strip}

\section{Limitations and Future Work}

The principal limitation follows directly from our efficiency choice.  A 65.32M-parameter model trained from scratch on 13,450 images possesses far less semantic knowledge than a pretrained text-to-image system: it learns JPEG blocks, edges, gradients, and recurring local textures, but cannot reliably infer object-specific detail once compression has removed the supporting evidence.  The gap at QF 5 and on MANIQA and DISTS quantifies this weakness, and outputs should therefore not be interpreted as faithful recovery of information absent from the bitstream.

Training scale is likewise modest: the reported model was trained on a single RTX 3090 for roughly 101k updates on a small fraction of LSDIR.  A natural next step is distributed training on the full LSDIR corpus, with crop resolution increased progressively throughout training rather than in a short final continuation.  Such a study would clarify whether the remaining quality gap is primarily data- and compute-limited or intrinsic to the absence of a pretrained prior.

Larger models and longer large-crop training may further improve texture knowledge while preserving the VAE-free pipeline.  In particular, a progressive-resolution schedule, in which the input resolution is increased in steps over the course of training, may improve generalization to the large images encountered at evaluation.

Finally, the cross-paper efficiency comparison inherits differences in software stacks even when input and batch sizes are aligned, and our architecture scaling study, though controlled on a single GPU, measures isolated forward passes rather than trained quality for every backbone.  Future work should report end-to-end latency, energy, and peak reserved memory under a single public harness across all methods and accelerators.

\section{Conclusion}

We have presented a compact generative JPEG restoration model designed around the quality-efficiency frontier.  An exact two-level Haar transform removes the need for a learned VAE, a conditional residual formulation preserves observed content, and a RELA-based linear DiT provides global interaction alongside a local full-rank path.  Clean-residual prediction parameterizes an improved MeanFlow objective, enabling two-update inference without distillation from a pretrained generator.  The model is not uniformly best in perceptual quality, since large semantic priors retain their advantage under severe compression, but its LPIPS, 65.32M parameters, and 8.05 images/s throughput at $1024^2$ demonstrate that high-throughput generative restoration is feasible without carrying a full text-to-image latent pipeline.

\section*{Acknowledgments}
We thank Radu Miron and Andrei Ciortea for their valuable feedback and helpful discussions. 

\begingroup
\small
\setlength{\bibsep}{2pt plus 0.3pt}
\bibliography{main}
\endgroup

\clearpage
\onecolumn
\appendix
\section{Qualitative Results}

Generated predictions for $512\times512$ LSDIR crops.  Rows 1-3 use QF 10 and rows 4-6 use QF 5, left and right groups use 2 and 1 steps.

\begin{figure}[H]
\centering
\captionsetup{font=small,skip=2pt}
\setlength{\tabcolsep}{1.5pt}
\newcommand{\qualimg}[1]{\includegraphics[width=0.145\textwidth]{#1}}
\begin{tabular}{@{}ccc@{\hspace{7pt}}ccc@{}}
\textbf{JPEG} & \textbf{Ours} & \textbf{Reference} &
\textbf{JPEG} & \textbf{Ours} & \textbf{Reference}\\[2pt]
\qualimg{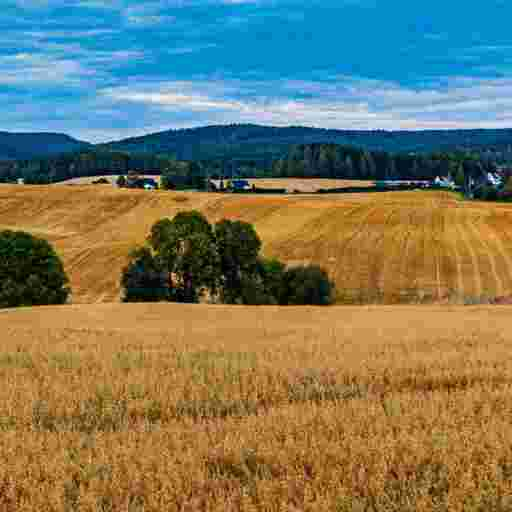} &
\qualimg{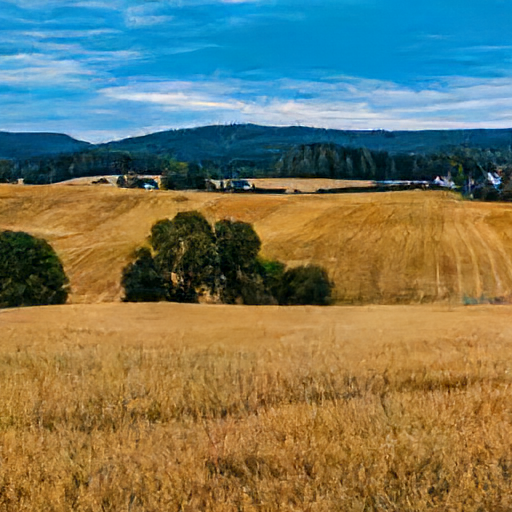} &
\qualimg{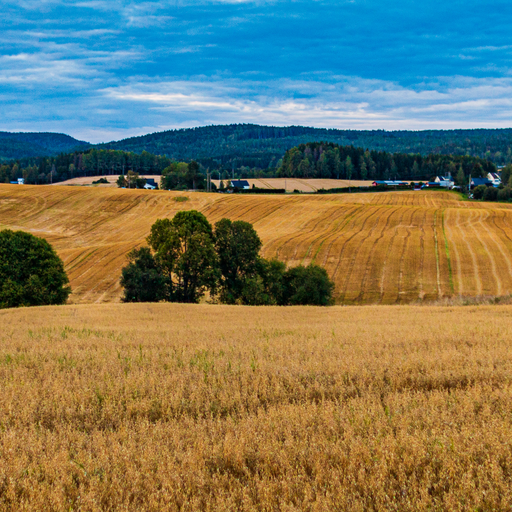} &
\qualimg{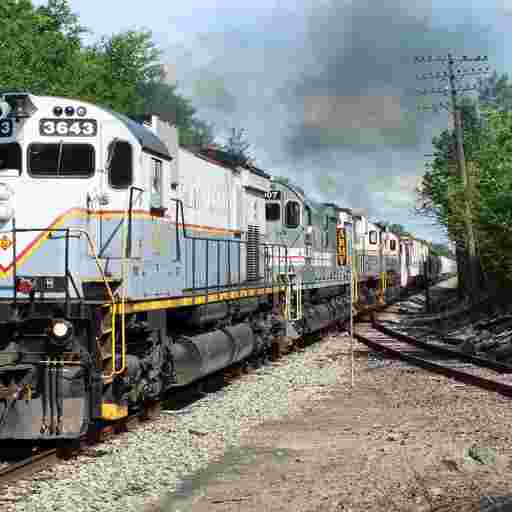} &
\qualimg{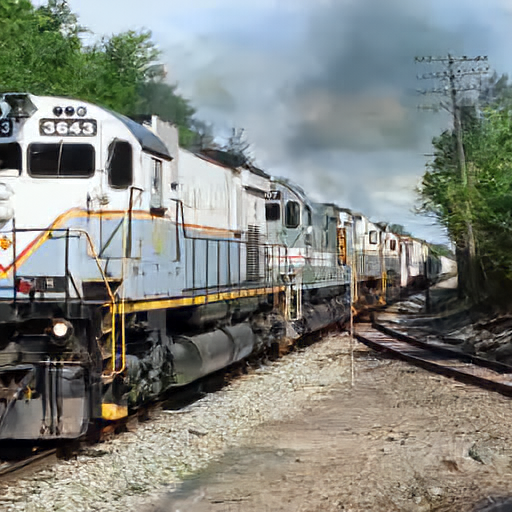} &
\qualimg{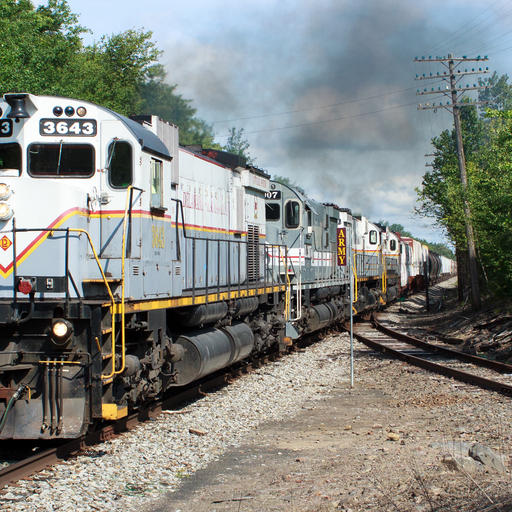}\\[-1pt]
\multicolumn{3}{c}{\footnotesize QF 10, 2 steps} & \multicolumn{3}{c}{\footnotesize QF 10, 1 step}\\[1pt]
\qualimg{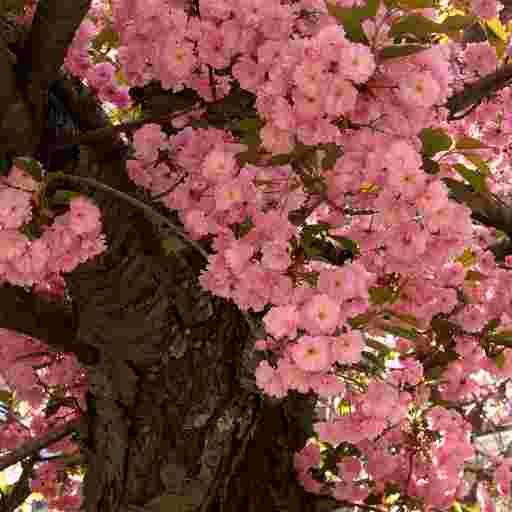} &
\qualimg{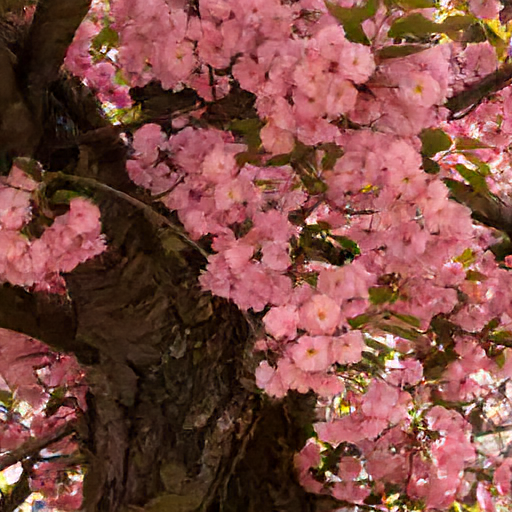} &
\qualimg{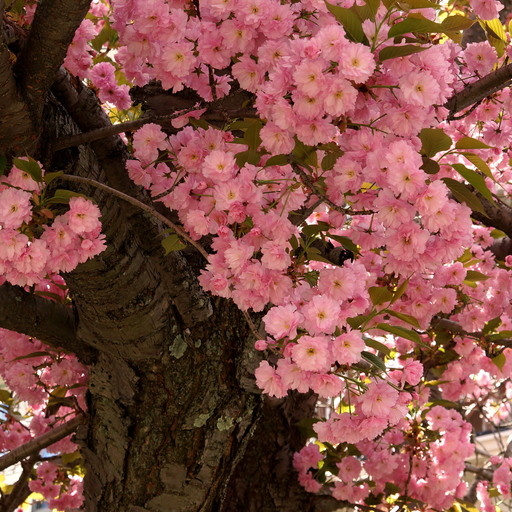} &
\qualimg{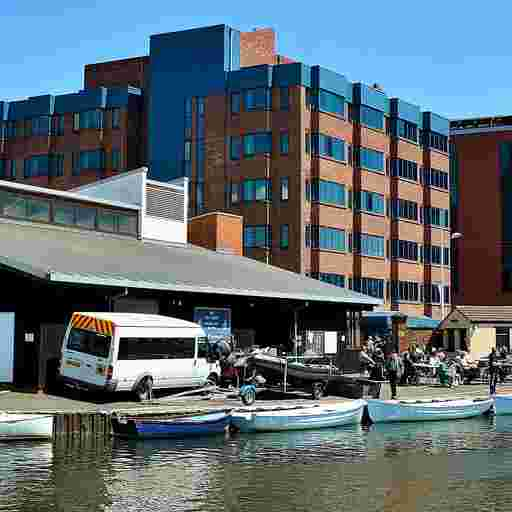} &
\qualimg{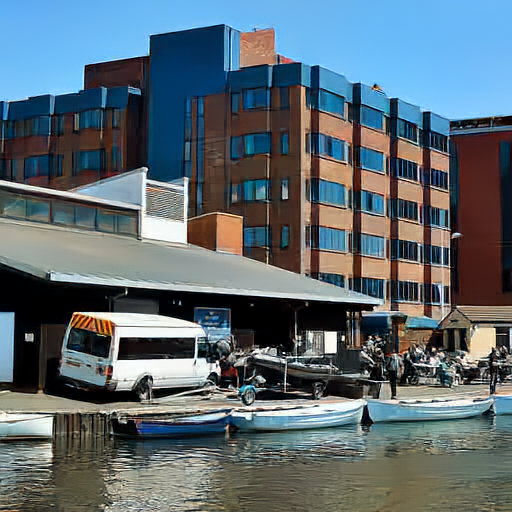} &
\qualimg{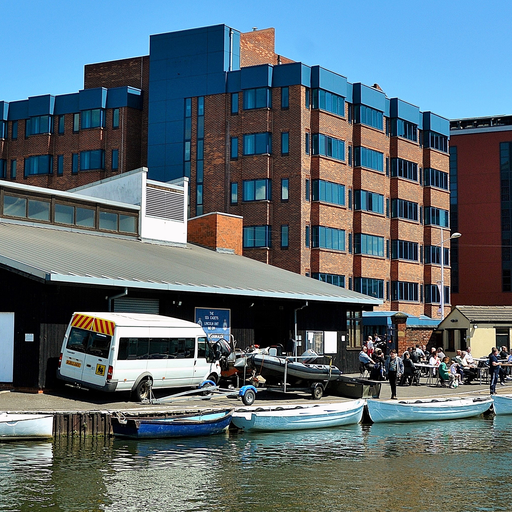}\\[-1pt]
\multicolumn{3}{c}{\footnotesize QF 10, 2 steps} & \multicolumn{3}{c}{\footnotesize QF 10, 1 step}\\[1pt]
\qualimg{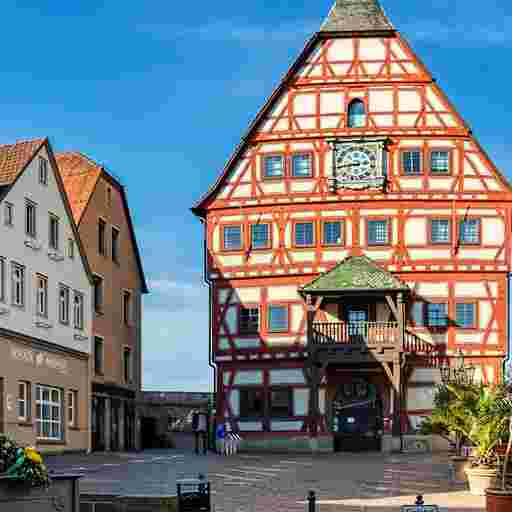} &
\qualimg{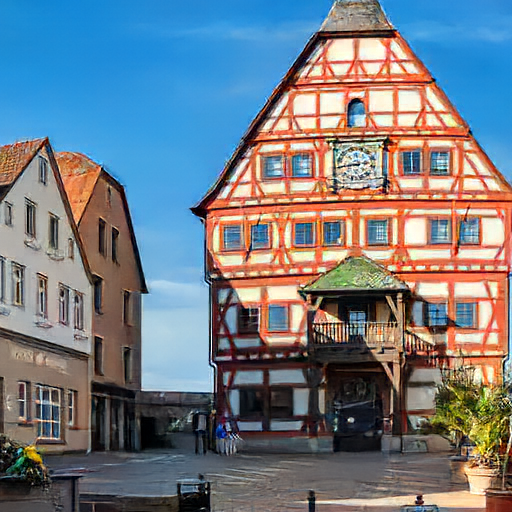} &
\qualimg{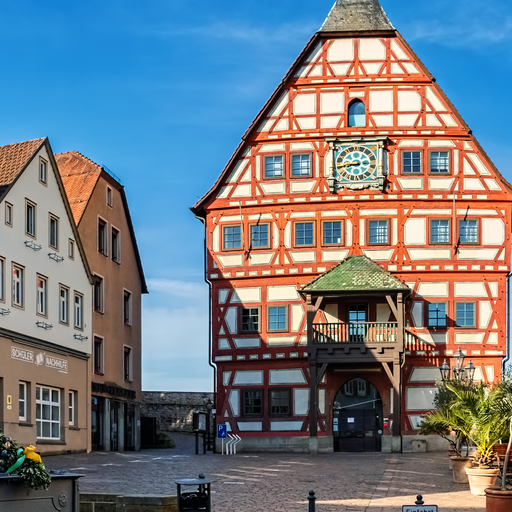} &
\qualimg{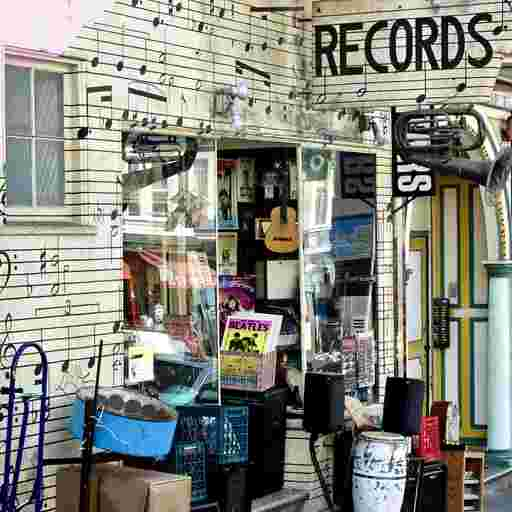} &
\qualimg{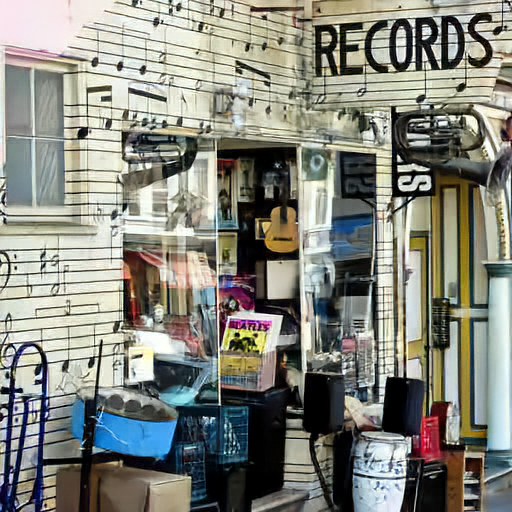} &
\qualimg{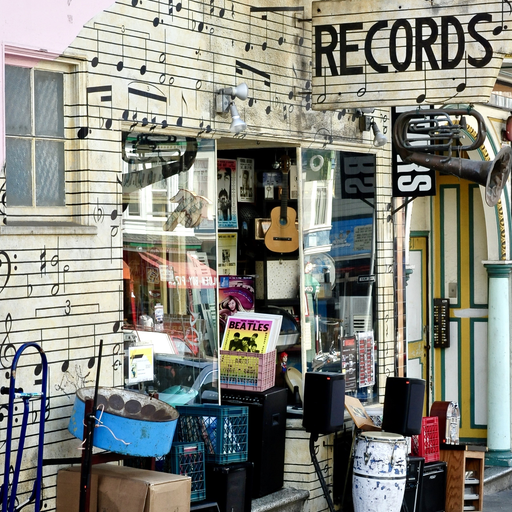}\\[-1pt]
\multicolumn{3}{c}{\footnotesize QF 10, 2 steps} & \multicolumn{3}{c}{\footnotesize QF 10, 1 step}\\[1pt]
\qualimg{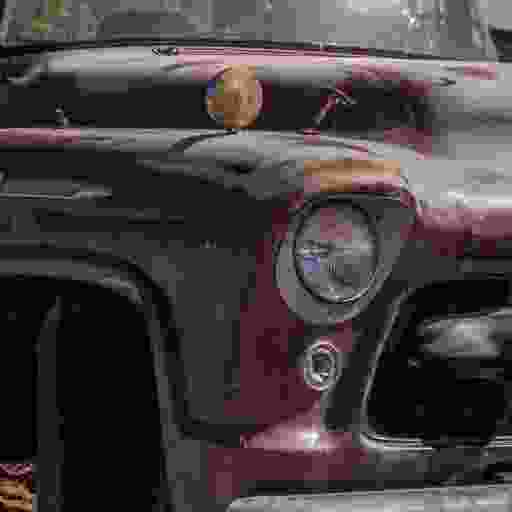} &
\qualimg{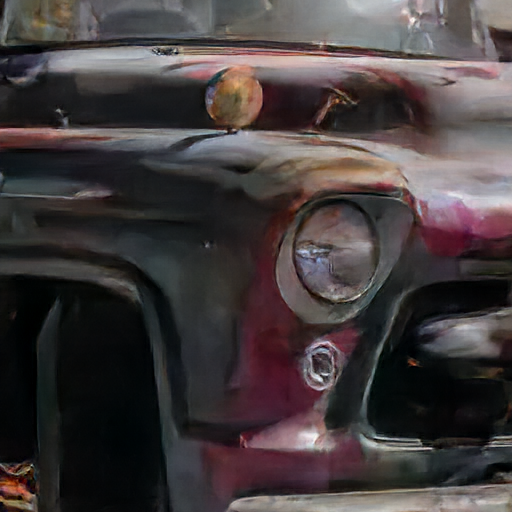} &
\qualimg{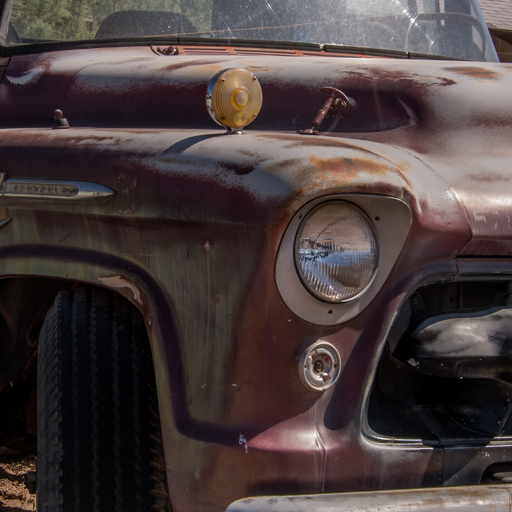} &
\qualimg{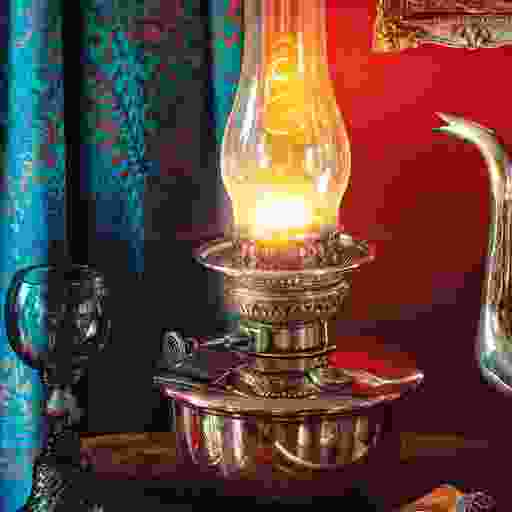} &
\qualimg{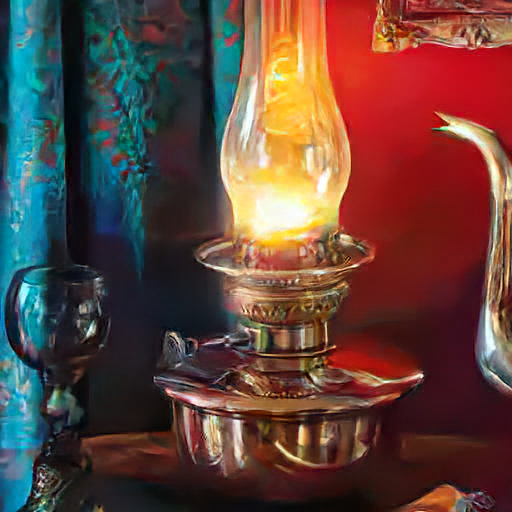} &
\qualimg{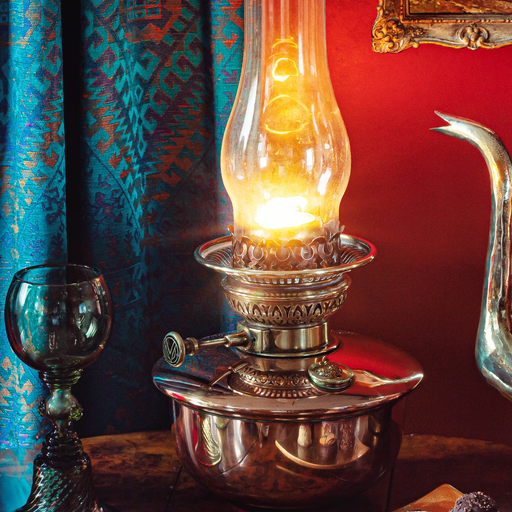}\\[-1pt]
\multicolumn{3}{c}{\footnotesize QF 5, 2 steps} & \multicolumn{3}{c}{\footnotesize QF 5, 1 step}\\[1pt]
\qualimg{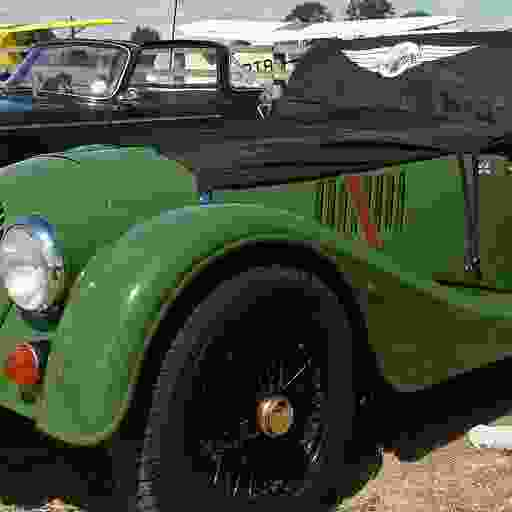} &
\qualimg{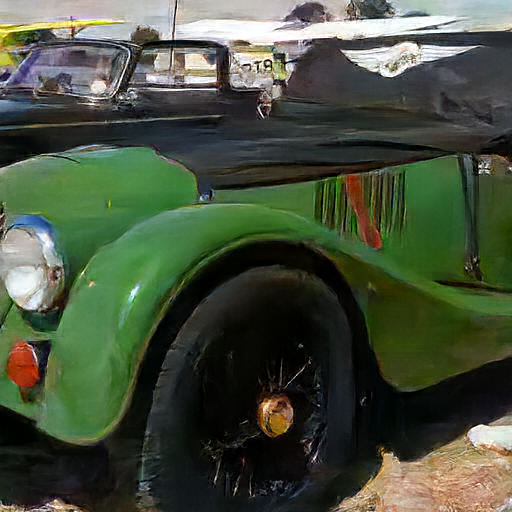} &
\qualimg{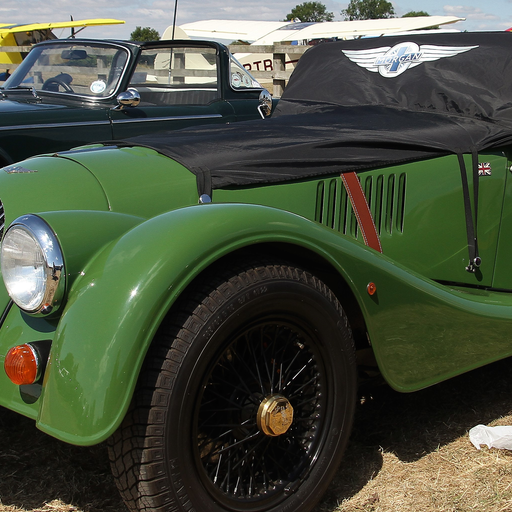} &
\qualimg{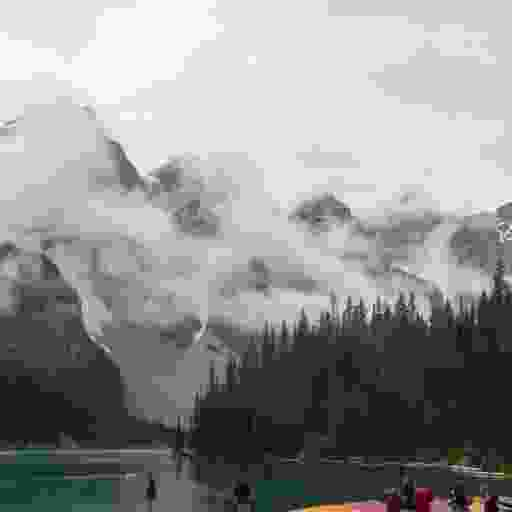} &
\qualimg{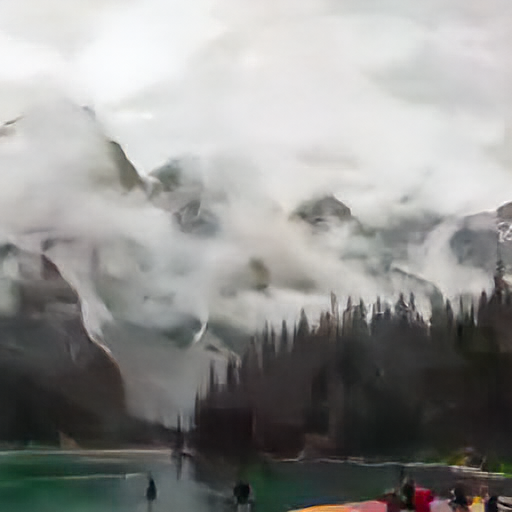} &
\qualimg{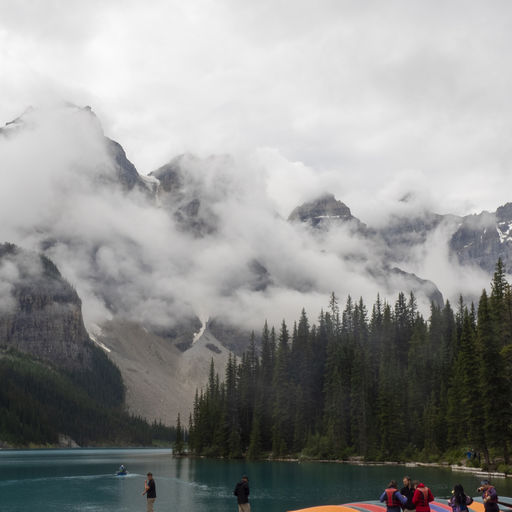}\\[-1pt]
\multicolumn{3}{c}{\footnotesize QF 5, 2 steps} & \multicolumn{3}{c}{\footnotesize QF 5, 1 step}\\[1pt]
\qualimg{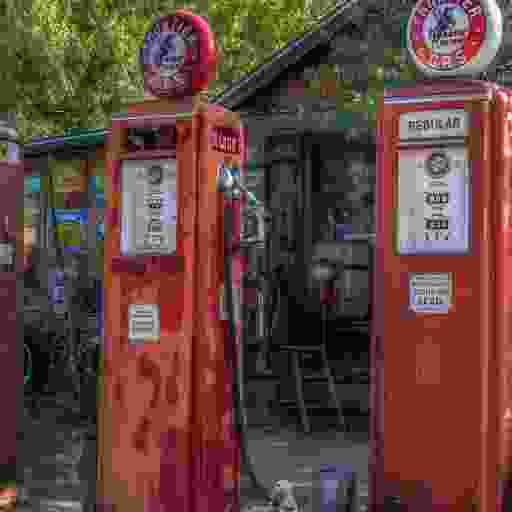} &
\qualimg{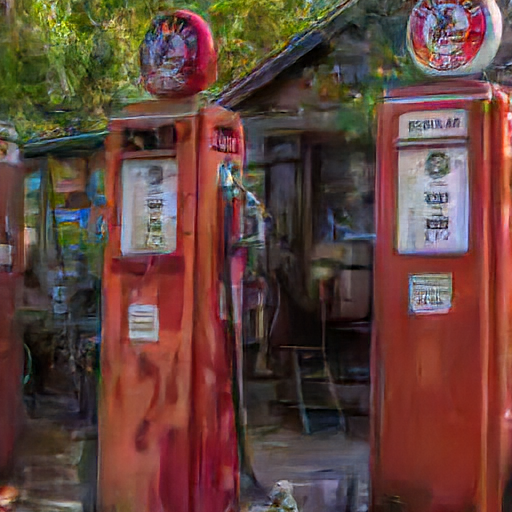} &
\qualimg{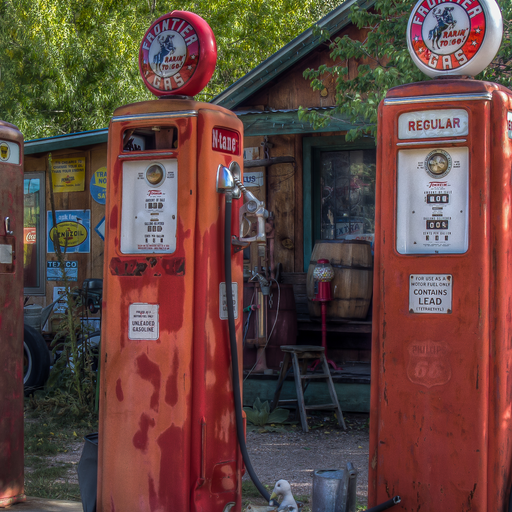} &
\qualimg{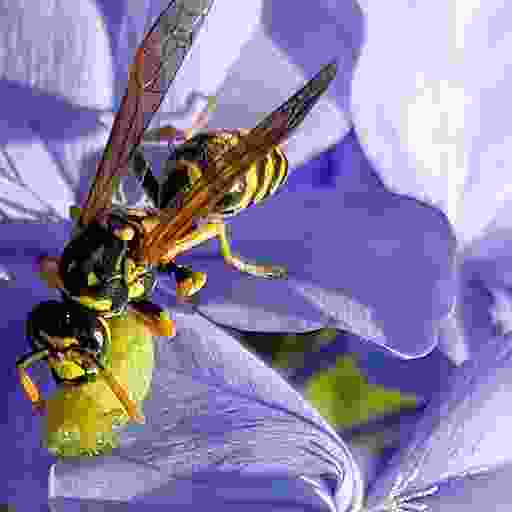} &
\qualimg{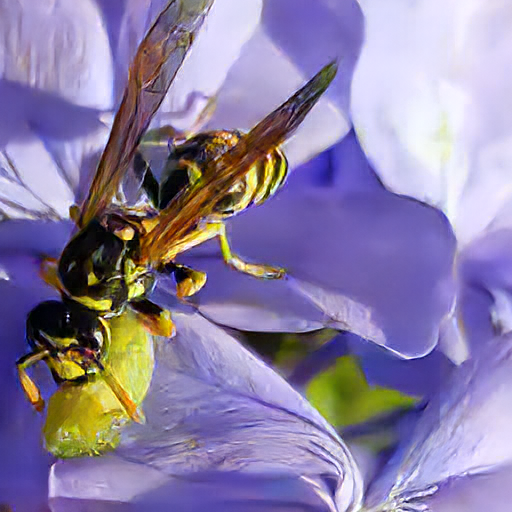} &
\qualimg{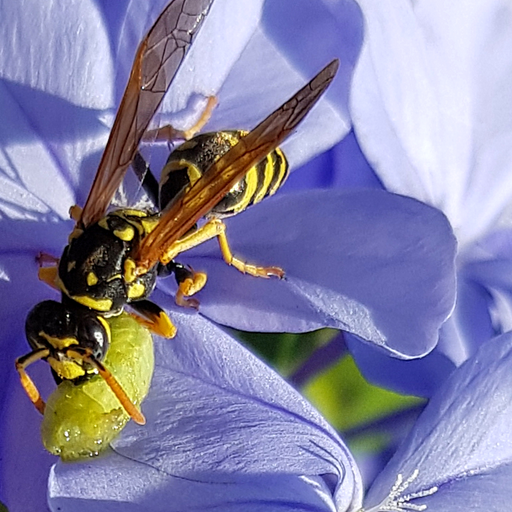}\\[-1pt]
\multicolumn{3}{c}{\footnotesize QF 5, 2 steps} & \multicolumn{3}{c}{\footnotesize QF 5, 1 step}
\end{tabular}
\end{figure}

\end{document}